\documentclass{article}

\usepackage{arxiv}
\usepackage{mathtools}
\usepackage[utf8]{inputenc} 
\usepackage[T1]{fontenc}    
\usepackage{hyperref}       
\usepackage{url}            
\usepackage{booktabs}       
\usepackage{amsfonts}       
\usepackage{nicefrac}       
\usepackage{microtype}      
\usepackage{lipsum}		
\usepackage{graphicx}
\usepackage{natbib}
\usepackage{doi}
\usepackage{amsmath}
\usepackage{amssymb}
\usepackage{tabularx, booktabs} 
\usepackage{multicol}
\usepackage{multirow}
\usepackage{subcaption}
\usepackage{float}
\usepackage[linesnumbered,ruled,vlined]{algorithm2e}
\usepackage{algorithmic}
\usepackage{caption}
\title{Task-Aware Few-Shot-Learning-Based Siamese Neural Network for Classifying Control-Flow-Obfuscated Malware}

\author{
  Jinting Zhu, Julian Jang-Jaccard, Amardeep Singh\\
  Cybersecurity Lab \\
  Massey University  \\
  Auckland, New Zealand\\
  \texttt{\{Jinting Zhu\}jzhu3@massey.ac.nz} \\
   \And
  Paul A. Watters \\
 Cyberstronomy Pty Ltd \\
 Melbourne, Australia\\
 \\
     \And
 Seyit Camtepe \\
  ‡CSIRO Data61, \\
 Australia\\
 \\
}

\renewcommand{\shorttitle}{\textit{arXiv} Template}

\hypersetup{
pdftitle={A template for the arxiv style},
pdfsubject={q-bio.NC, q-bio.QM},
pdfauthor={David S.~Hippocampus, Elias D.~Striatum},
pdfkeywords={First keyword, Second keyword, More},
}

\begin{document}
\maketitle

\begin{abstract}
Malware authors apply different techniques of control flow obfuscation, in order to create new malware variants to avoid detection. Existing Siamese neural network (SNN)-based malware detection methods fail to correctly classify different malware families when such obfuscated malware samples are present in the training dataset, resulting in high false-positive rates. To address this issue, we propose a novel task-aware few-shot-learning-based Siamese Neural Network that is resilient against the presence of malware variants affected by such control flow obfuscation techniques. Using the average entropy features of each malware family as inputs, in addition to the image features, our model generates the parameters for the feature layers, to more accurately adjust the feature embedding for different malware families, each of which has obfuscated malware variants. In addition, our proposed method can classify malware classes, even if there are only one or a few training samples available. Our model utilizes few-shot learning with the extracted features of a pre-trained network (e.g., VGG-16), to avoid the bias typically associated with a model trained with a limited number of training samples. Our proposed approach is highly effective in recognizing unique malware signatures, thus correctly classifying malware samples that belong to the same malware family, even in the presence of obfuscated malware variants. Our experimental results, validated by N-way on N-shot learning, show that our model is highly effective in classification accuracy, exceeding a rate \textgreater 91\%, compared to other similar methods.
\end{abstract}

\keywords{Siamese neural network \and meta-learning  \and malware classification  \and code obfuscation  \and few-shot learning}

\section{Introduction}
Malware producers are ever more motivated to create new variants of malware, in order to gain profits from unauthorized information stealth. According to the malware detection agency AV Test (\url{https://www.av-test.org/en/statistics/malware/}  (accessed on 1 July 2021)), 
100 million new variants of malware were generated from January to October 2020, which translates as roughly three thousand new malware daily. 

In particular, we have witnessed the fast growth of mobile-based malware. An NTSC report published in 2020  \url{https://www.ntsc.org/assets/pdfs/cyber-security-report-2020.pdf}  (accessed on 15 June 2021) reported that 27\% of organizations globally have been impacted by malware attacks sent via Android mobile devices. In recent times, we have seen malware producers employ techniques such as obfuscation \cite{dong2018understanding,chua2018effectiveness, bacci2018detection} and repackaging \cite{song2017appis,lee2019enhanced,zheng2017security}, mostly through the change of static features \cite{zhu2018droiddet,sun2017malware,hu2014encoding}, to avoid detection. 
In response to the trend in the growth of mobile-based malware attacks, numerous Artificial Intelligence (AI)-based defense techniques have been proposed \cite{vasan2020image, luo2017binary, su2018lightweight,makandar2018trojan, hsiao2019malware, singh2019migan}. 

We argue that there are two main issues to be addressed in the existing state-of-the-art of AI-based mobile malware attack defense. 

The first issue is that most of the existing research tends to focus on learning from common semantic information about the generic features of malware families, and building feature embeddings \cite{raff2018investigation, gibert2019hierarchical, shen2018feature}. In this context, 'feature embedding' means the features contained in a malware binary sample, which provide important clues as to whether the malware image---generated from the hexdump utility that displays the contents of binary files in hexadecimal, from which feature embedding is created---is malicious or not: if malicious, what type of malware family it belongs to is assessed, to build the right set of response strategies. These existing works often treat the fraction of the code changed by the obfuscation and repackaging as a type of noise \cite{gibert2018classification}, and thus tend to ignore the effect of the modification: this is largely because the code changed by the obfuscation and repackaging techniques displays a similar appearance when malware visualization techniques are applied \cite{akarsh2019detailed,ni2018malware,naeem2020malware}. Using common semantic information as data input points to be fed into a deep neural network cannot capture the unique characteristics of each malware family signature: thus, they will not be able to accurately classify many variants arising from the same malware family \cite{kalash2018malware, milosevic2017machine, vasan2020image, yuan2020byte}, especially if an obfuscation technique is applied. 

The second issue with the existing approaches is the demand for large data input, with which to find more relevant correlations across the features: such input is unable to detect and classify malware families trained with a limited number of samples (e.g., newly emerging variants of malware) \cite{cao2018softmax}.

To address these two important issues, we propose a novel task-aware few-shot-learning based Siamese neural network, capable of detecting obfuscated malware variants belonging to the same malware family, even if only a small number of training samples are~available.

The contributions of our proposed model are as follows:
\begin{itemize}


	\item Our task-aware meta-learner network combines entropy attributes with image-based features for malware detection and classification. By utilizing the VGG-16 network as part of the meta-learning process, the weight generator assigns the weights of the combined features, which avoids the potential issue of introducing bias when the training sample size is limited;
	
	
	
	\item For the hybrid loss to compute the intra-class variance, the center loss is added alongside the constructive loss, to enable positive pairs and negative pairs to form more distinct clusters across the pairs of images processed by two CNNs;

	\item The results of our extensive experiments show that our proposed model is highly effective in recognizing the presence of a unique malware signature, despite the presence of obfuscation techniques, and that its accuracy exceeds the performance of similar methods.
	
\end{itemize}

We organized the rest of the paper as follows. We examine the related work in Section~\ref{sec:rw}. We describe how control-flow-obfuscated malware variants are created, and we address why the generic SNN approach cannot detect such obfuscated malware variants in Section ~\ref{sec:preliminary}. We provide the details of our proposed model, along with the details of the main components, their roles and responsibilities, and an overview of the algorithm involved, in Section~\ref{sec:our_model}. In Section~\ref{sec:experiment}, we describe the details of the dataset, feature extraction, and the experimental results with analysis. Finally, we provide the conclusion to our work, including the limitations of our proposal, and future work directions, in Section~\ref{sec:conclusion}.

\section{Related Work}\label{sec:rw}

In this section,  we review two lines of research relevant to our study: few-shot-learning-based malware detection and feature embedding applied for malware detection. 
\subsection{Entropy Feature in Feature Selection}
Entropy value serves as a metric for feature selection, quantifying the information each feature contributes to the target variable or class. Selecting features with the highest information gain can enhance model performance and reduce overfitting. Huang et al. \cite{huang2016sparse} utilized the entropy function defined on sparse signal $x$ to recover such signals: minimizing it with an appropriate p-value yielded sparse solutions and improved signal recovery rates. Additionally, Finlayson et al. 
\cite{finlayson2009entropy} demonstrated that quadratic entropy values, being smoother and typically having a single minimum, offer the most efficient approach. This method aids full-color shadow removal, by comparing edges in the original and invariant images, followed by re-integration. Moreover, Allahverdyan et al. \cite{kolouri2018joint, allahverdyan2018adaptive} suggest that entropy minimization can lead to basic forms of intelligent behavior. Specifically, Kolouri et al. \cite{kolouri2018joint} employed entropy minimization to bolster classifier confidence, while entropy regularization ensured that predictions remained close to unseen attributes in zero-shot tasks.

\section{Preliminary}\label{sec:preliminary}

\subsection{Control Flow Obfuscation}
Malware obfuscation is a technique that is applied by malware authors to create new malware variants, in order to avoid detection without creating a completely brand-new malware signature. Among the many different obfuscation techniques, we focused on control flow obfuscation, which involves creating a new malware variant by reordering the control flow of functional logic from the original malware program \cite{dong2018understanding}. This type of obfuscation technique makes the compiled malicious code appear to be different from the existing malware signature so that it can easily avoid detection. Our model can detect three different types of control flow obfuscation.

\subsubsection{Function Logic Shuffling}

\begin{figure}[h]
\centering
\includegraphics[scale=0.6]{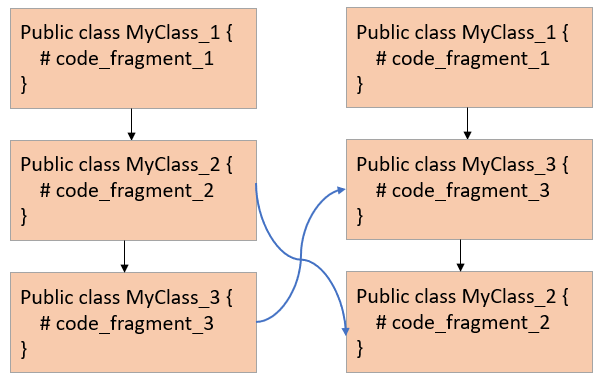}
\caption{Function Logic Shuffling.}
\label{fig:obf_ex1}
\end{figure}

This technique alters the control flow path of a malware program, by shuffling the order of function calls without affecting the semantics (i.e., purpose) of the original malware program: while the functionality between the original malware and the obfuscated version remains the same, the changing of appearance in the compiled code can result in the appearance of the malware image changing, and detection accuracy decreasing.  An example is shown in Figure \ref {fig:obf_ex1}, where the order of function logic MyClass\_2 and MyClass\_3 is changed.

\unskip

\subsubsection{Junk Code Insertion}

\begin{figure}[H]
	\centering
	\includegraphics[scale=0.59]{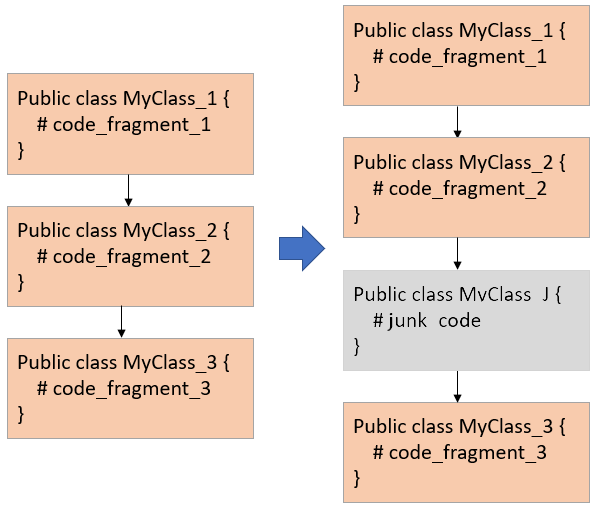}
	\caption{Junk Code Insertion.}
	\label{fig:obf_ex2}
\end{figure}

In this technique, the malware author inserts much (junk) code that never gets executed, whether after unconditional jumps, calls that never return, or conditional jumps with conditions that will never be met. The main goal of this technique is to change the control flow path, in order to avoid detection or to waste time for the reverse engineer analyzing useless code. An example of a junk code insertion is shown in Figure \ref {fig:obf_ex2}, where a junk code, MyClass\_J, is added in between two normal function calls.

\subsubsection{Function Splitting}
\begin{figure}[H]
		\centering
	\includegraphics[scale=0.7]{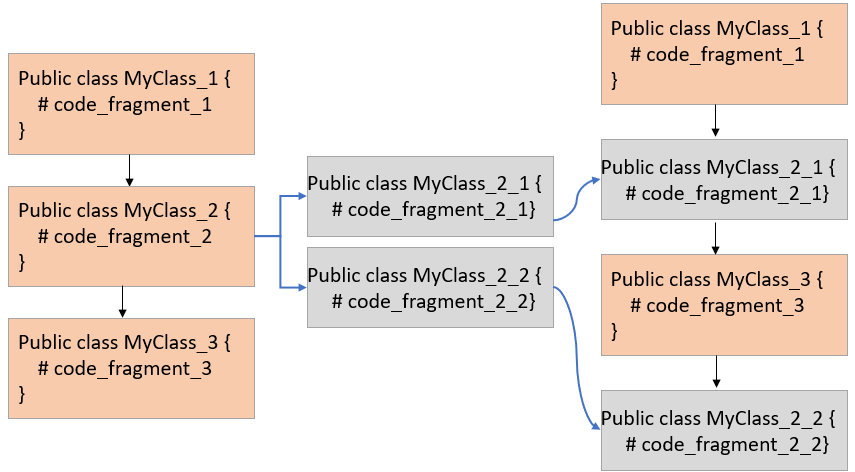}
	\caption{Function Splitting.}
	\label{fig:obf_ex3}
\end{figure}

With this technique, the malware author applies the function splitting method, where a function code is fragmented into several pieces, each of which is inserted into the control flow. This technique splits the function into $n$ code fragments, or merges pieces of unrelated codes, to make the changes in the compiled code, which also results in the malware image appearance changing. An example of a function splitting is shown in Figure \ref {fig:obf_ex3}, where two splits from MyClass\_2 are generated, and randomly added among other function calls.

Though the functionality between the original malware and obfuscated versions (e.g., malware variants) stays the same, a malware code applied with control flow obfuscation can easily avoid detection from many anti-virus programs \cite{dong2018understanding}. To address this issue, we propose a model resilient against the presence of many variants of malware created as a result of applying the control flow obfuscation technique. Our proposed method utilizes the information gain calculated through the entropy features associated with each malware variant. In our proposal, the entropy features measure the amount of uncertainty of a given probability distribution of a malware program that is not affected by the order of functional logic of the malware program.

\subsection{Generic Approach and Issues}
\begin{table}[H]
	
	\caption{Notations in task-aware meta learning-based Siamese neural network.\label{tab1}}
	
	\newcolumntype{C}{>{\centering\arraybackslash}X}
	
	\begin{tabularx}{\textwidth}{cl}
		\toprule
		\textbf{Notation}	& \textbf{Description}	\\
		\midrule
		$f_p$  &The log probability calculated by the $d_w^{(i)}$.  \\
		$f_q$  &The log probability calculated by the $1-d_w^{(i)}$. \\
		$\mathbf{x_i}$  &The malware image feature of $i$th samples.\\
		$y_t^i$ & The class $t$ of $i$th samples. \\ 
		$\theta_{t}$  & The feature layers' parameters.  \\ 
		$\mathbf{W^i}$  & I-th feature layer in $\mathcal{F}$'s weights. \\ \
		$\mathbf{W^i_{sr}}$  & The shared parameters for all malware families. \\ 
		$\mathbf{W^i_{ts}}$  & The task-specific parameters for each malware family. \\ 
		$\mathbf{c_i}$ & The center point of each class. \\ 
		$L_e$ & The embedding loss.  \\ 
		$L_{b,c}$	& The binary cross entropy loss with center loss.  \\	
		$\mathbf{d_w^i}$ & The distance feature of a pair of images.	\\	
		$y_d$   & The label of pairs of images. \\	
		$\mathbf{F_w}$   & The convolutional filter with parameters. $w$ 	\\ 
		$\beta$ & The hyper-parameter.  \\ 
		\bottomrule
	\end{tabularx}
\end{table}
In the last few years, few-shot learning technology, for example, the Siamese neural network (SNN), which uses only a few training samples to get better predictions has emerged. An SNN contains two identical subnetworks (usually convolutional neural networks)---hence the name 'Siamese'. The two CNNs have the same configuration, with the same weights, $\textbf{\textit{W}} \in \mathbf{R}^d$, where $\textbf{\textit{W}}$ depicts the model's parameters, while $\textbf{R}^d$ depicts the distance embedding to calculate two samples inputted from each subnetwork, respectively, and the value of the distance indicates whether they are closed to each other, as the value increases or decreases in the Euclidean space. The updating of the hyperparameters is mirrored across both CNNs, and is used to find the similarity of the inputs, by comparing its feature vectors.

Each parallel CNN is designed to produce an embedding (i.e., a reduced dimensional representation) of the input. These embeddings can then be used to optimize a loss function during the training phase and to generate a similarity score during the testing phase.

The architecture of Spiking Neural Networks (SNNs) contrasts significantly with traditional neural networks, which rely on extensive datasets to learn the prediction of multiple classes. The latter's requirement for total retraining and updating with each addition or removal of a class presents a challenge. SNNs instead learn through a similarity function, testing whether two images are identical. This innovative architecture empowers them to classify new data classes without the need for additional network training. Furthermore, SNNs demonstrate greater robustness against class imbalance, as a small number of images per class is enough to enable future recognition. The corresponding notations are listed as Table \ref{tab1}.

\begin{figure}[H]
	\centering
	\includegraphics[scale=0.45]{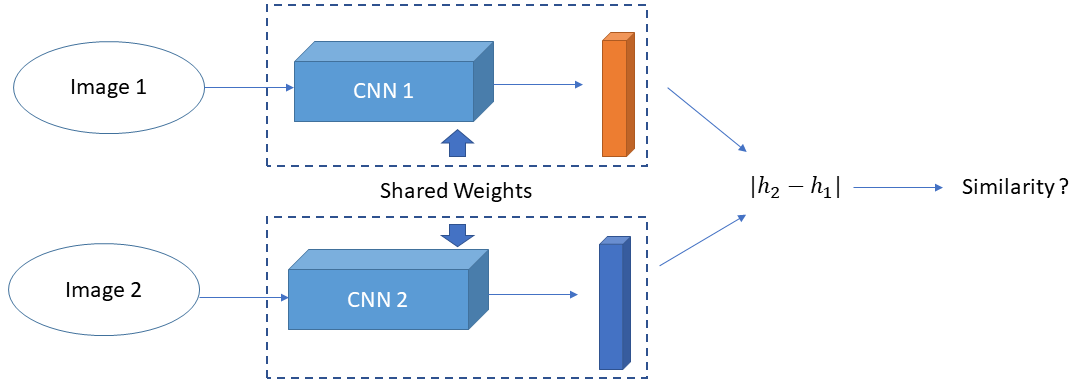}
	\caption{Overview of a Siamese neural network.}
	\label{fig:snngeneric}
\end{figure}

Figure \ref{fig:snngeneric} illustrates the working of a generic SNN, its goal being to determine if two image samples belong to the same class or not: this is achieved through the use of two parallel CNNs (CNN1 and CNN2 in Figure \ref{fig:snngeneric}) trained on the two image samples (Image 1 and Image 2 in Figure \ref{fig:snngeneric}). Each image is fed through one branch of the SNN, which generates a d-dimensional feature embedding ($h_1$ and $h_2$ in Figure \ref{fig:snngeneric}) for the image: it is these feature embeddings that are used to optimize a loss function, rather than the images themselves. A supervised cross-entropy function is used in the SNN for the binary classification, to determine whether two images are similar or dissimilar, by computing $[h_2$-$h_1]$ and processing it by a sigmoid function. 
Mathematically, the similarity between a pair of images ($x_1, x_2$) within Euclidean distance (ED) is computed in an SNN using the following equation:
\begin{equation}
	\mathbf{d_w} (\mathbf{x_1},\mathbf{x_2})=\|F_w (\mathbf{x_1})-F_w (\mathbf{x_2} ) \|_2 ,
\end{equation}
where the $F_w$ indicates the feature representation for the inputted feature matrix. Generally, the model $g_w$: $ \mathbf{R}^n$ to $\mathbf{R}^d$ is  parameterized by the weights $\textbf{\textit{w}}$, and its loss function is calculated using the following equation:
\begin{equation}
	L_b=-\frac{1}{N} \sum_{i=1}^{N}[\mathbf{y^{i}_{d}}f_{p}(\mathbf{d_{w}^{(i)}})+(1-\mathbf{y_d^i})f_{q}(\mathbf{d_{w}^{(i)}})],
\end{equation}{}
where $\mathbf{y_d} = \{\mathbf{y_d^1, y_d^2, \ldots , y_d^i } \}\in\{0, 1 \}$ denotes the ground truth label of image pairs ($\mathbf{x_i,x_j}$), and $d_w$ represents the Euclidean distance (ED) between two images at the $i$-th pair. Note that the most similar images are supposed to be the closest in the feature embedding space:  though this approach would work well for finding similarities/dissimilarities across distinct image objects, it would not work well for obfuscated malware samples. 

Recall that an obfuscated malware---for example, $\mathbf{x_1}$---changes some part of the original malware code, $x_2$: when these two are converted as a feature representation---for example, $F_w (\mathbf{x_1)}$ and $F_w (\mathbf{x_2})$---the feature values in the feature representation will look very different, which is how obfuscated malware avoids detection by anti-virus software. Inadvertently, the different values in the feature representation make the distance across obfuscated malware images very different from one another (i.e., $d_w (\mathbf{x_1,x_2})$ is large). Eventually, when a similarity score is computed and compared, using the loss (i.e., $L_b$) based on the distance calculation (i.e., $d_w (\mathbf{x_1,x_2})$), they appear to be different malware families---though, in fact, they all belong to the same malware family.

\section{Task-Aware Meta Learning-Based Siamese Neural Network} \label{sec:our_model}

We now introduce our task-aware meta-learning-based SNN model, which provides a novel feature embedding better-suited to control-flow-obfuscated malware classification. We start with the overview of our proposed model, and the details of the CNN architecture that is used by our model, followed by how task-specific weights are calculated using factorization. Finally, we discuss the details of the loss functions our model uses to address the challenges of weight generation with a limited number of training samples.

\subsection{Our Model}

\begin{figure}[H]
	\includegraphics[scale=0.49]{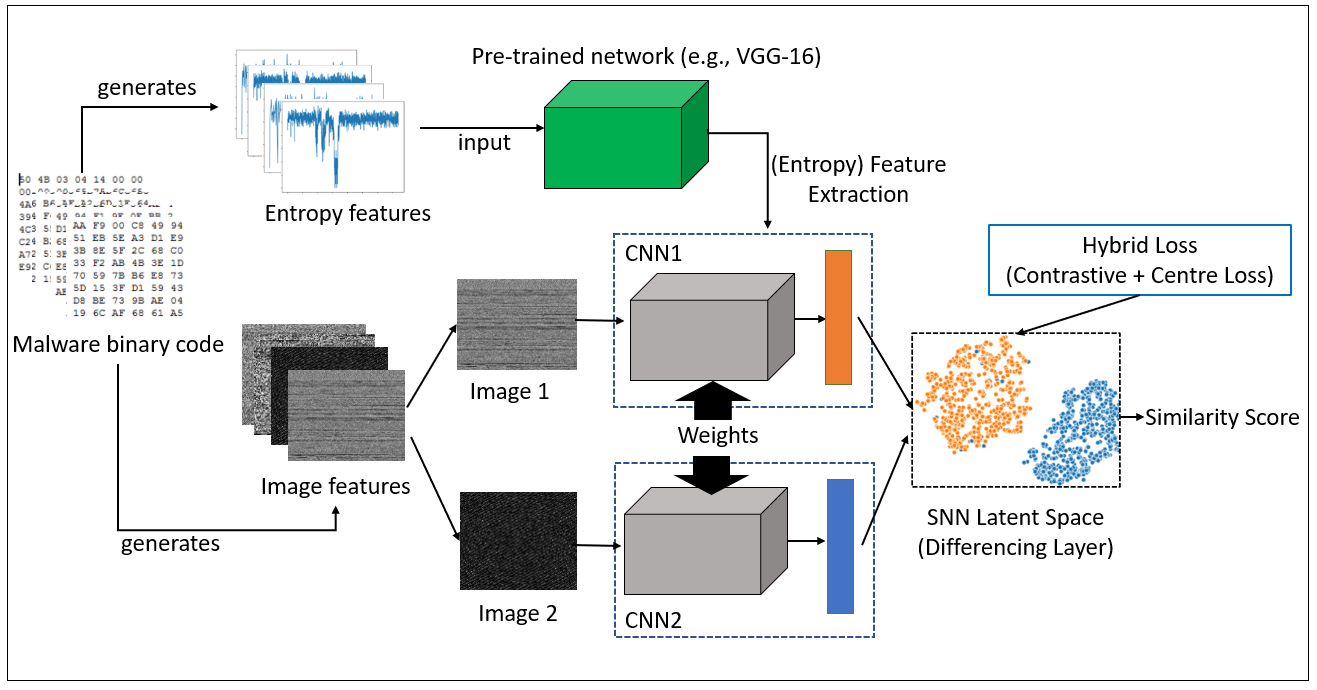}
	\caption{Overview of our proposed model.}
	\label{fig:task-ware}
\end{figure}

As shown in Figure \ref{fig:task-ware}, our model utilizes a pre-trained network and two identical CNN networks. We use a pre-trained network (VGG-16) to compute more accurate weights for entropy features. Similarly, each CNN takes image features to calculate weight for the image features, and to generate feature embedding using the task-specific weights and shared weights of both the entropy and the image features. The feature embeddings produced by the two CNNs are used by the SNN to calculate the similarity score across intra-class variants, using a new hybrid loss function.

Within each CNN, there are two sub-networks: a task-aware meta-learner network and an image network, as shown in Figure \ref{fig:cnn_architecture}. The task-aware meta-learner network starts by taking a task specification (e.g., an entropy feature vector), and generates the weights. At the same time, the image network (e.g., a typical CNN branch of an SNN) starts by taking the image feature and convoluting it, until a fully connected layer is produced. The last fully connected layers use the weights generated by the task-aware meta-learner network, along with the shared weights, to produce a task-aware feature embedding space. Embedding loss for inter-class variance is calculated for back-propagation, until the CNN is fully trained for all input images.

Mathematically, this can be written as the following equation:
\begin{equation}
	y = \mathcal{F}(x; \theta = \{\mathcal{G}(t), \theta_{f}\}) ,
\end{equation}
where the SNN $\mathcal{F}$ takes malware images $\mathbf{x}$ as inputs, and produces a task-aware feature embedding that is used by the SNN to predict the similarity $\mathbf{y} \in \{1,0\}$ between an image pair inputted to two CNNs. Each CNN is parameterized by the weights $\theta$, which are composed of generated parameters from $\mathcal{T}$ and the share parameters $\theta_{s}$ in the SNN $\mathcal{F}$ that is shared across all malware families. The task-aware meta-learner network $\mathcal{T}$ creates a set of weight generators $g^i, i = {1\ldots k}$, to generate parameters for $k$ feature layers in $\mathcal{F}$, conditioned on $e_t$. 
The overall approach of our proposed model can also be summarized using Algorithm \ref{alg:alg1}:

\begin{figure}[H]
	\includegraphics[scale=0.6]{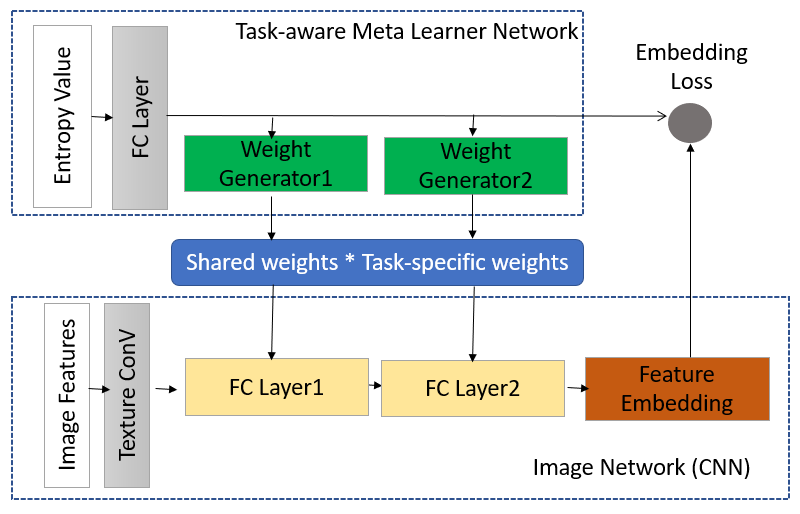}
	\caption{CNN architecture of our proposed model.}
	\label{fig:cnn_architecture}
\end{figure}
\vspace{-12PT}

\begin{algorithm}
	\caption{Pseudo-code of our proposed algorithm.}
	\label{alg:alg1}
	Binary cross-entropy with center loss:$L_{b, c}$\;
	Additional supervision loss :$L_e$\;
	\SetKwInOut{Input}{Input}
	\SetKwInOut{Output}{Output}
	
	\Input{Entropy graph feature $F_{ent}$, texture feature $F_t$,  support set $s$, query set $q$, pair label $y_d^i$, sample label $\mathbf{y_t^i}$, hyper-parameters $\beta$ , initialized centers $c_i$}
	\Output{[Predicted similarity]}
	
	\textbf{Training stage:}
	
	(1) Initializing the parameters for our proposed models, and the task-specific weights $W^i$ for the weight generators in the task-aware meta-learner network $g^i$, using the weight factorization Equation \eqref{eq:factorazation};
	
	(2) To input the malware texture features $F_t$ and the 4096 features of entropy value $F_{ent}$ extracted by the pre-trained network (e.g., VGG-16); note that these are integrated with the support set $s$ of our proposed model;
	
	(3) The weighted features from Equation (5) are fed into the embedding loss, according to the one-shot label generated from the target label;\\
	
	(4) Calculating the Euclidean distance (ED) of features in the two branches of SNN, through the hybrid loss function; \\
	(5) Back-propagation and update parameters by Adam optimizer.\\ 
	
	\textbf{Testing Stage:}
	
	\While{not reach to iterations}{
		Extract features of samples in the query set $q$, and feed them the one-shot accuracy.}
	Accuracy = 100$\times correct/iterations\;$
	
\end{algorithm}
\vspace{-12PT}

\subsection{Task-Aware Meta-Learner}
Our task-aware meta-learner provides two important functionalities: one is generating optimized task-specific weights, using the entropy values extracted from a pre-trained deep learning model (e.g., VGG-16); 
the other function is to work with the image network, to compute the new weights, based on the shared weights and the task-specific weights, so that the embedding loss is accurately calculated, in order to capture the relative distance across inter-class variance (e.g., the features of the image). These functions are necessary, because some malware samples (e.g., zero-day attack samples) are usually much smaller than the number of images required for training an SNN model.

	
Using the entropy values, our meta-learner recognizes a specific malware signature present in the entropy, so that later it uses this knowledge to find whether some malware samples are derived from the same malware family or not (e.g., obfuscated malware). The entropy values are extracted from the VGG-16 when entropy graphs are inputted. 
	
\begin{figure}[H]
		\subfloat[\centering Sample 1 of Wroba.wm class]{%
			\includegraphics[width=.45\textwidth]{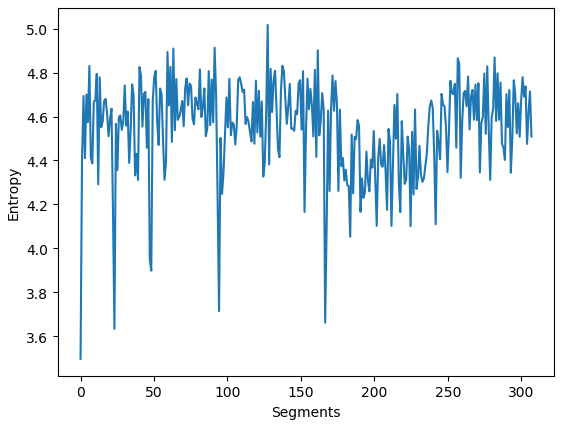}%
			\label{fig:y equals x}%
		}\hfill
		\subfloat[\centering Sample 2 of Wroba.wm class]{%
			\includegraphics[width=.45\textwidth]{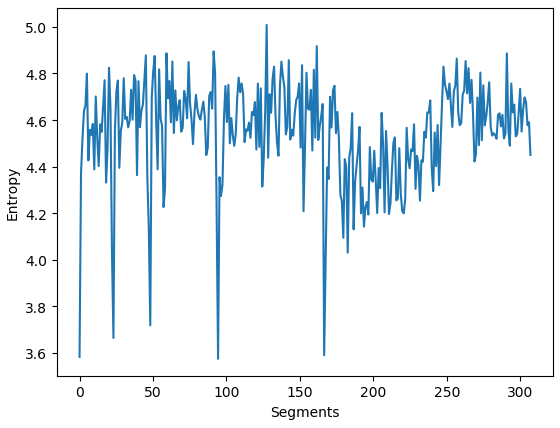}%
			\label{fig:y equals x}%
		}\hfil
		\subfloat[\centering Sample 1 of Agent.ad class]{%
			\includegraphics[width=.45\textwidth]{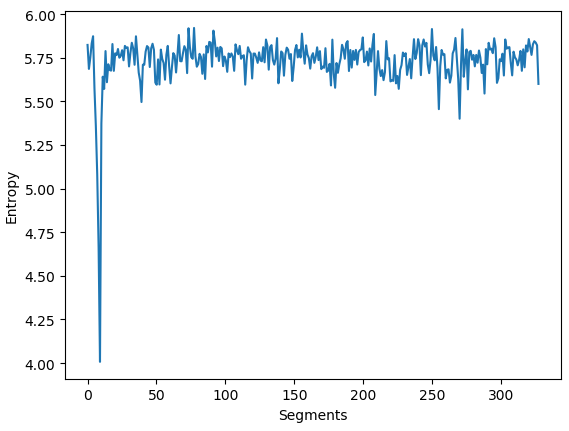}%
			\label{agent.ad.png}%
		}\hfill
		\subfloat[\centering Sample 2 of Agent.ad class]{%
			\includegraphics[width=.45\textwidth]{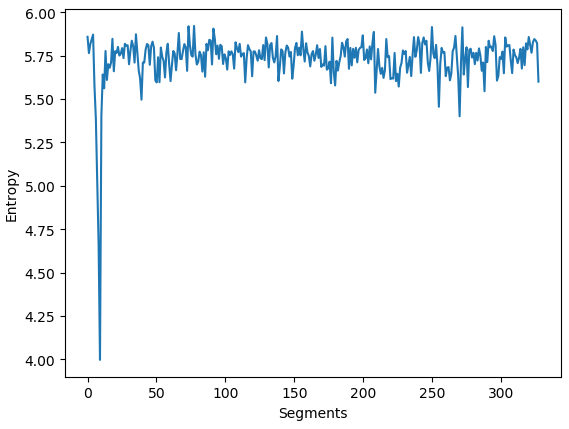}%
			\label{fig:y equals x}%
		}
		\caption{Examples of entropy graphs of two malware families.}
		\label{fig:entropy}
\end{figure}

We use entropy graphs to recognize a unique malware signature belonging to each malware family. To illustrate the use of an entropy graph as a task specification, four samples of malware images are shown in Figure \ref{fig:entropy}. Figure \ref{fig:entropy}a,b are two obfuscated malware samples from the same Wroba.wm family. Similarly, Figure \ref{fig:entropy}c,d of the name Agent.ad are from the same Agent family. One can see that the entropy graphs within the same family share a similar pattern, while there are visible differences in the entropy graphs between two different malware families.
\vspace{-9pt}
	
Our task-aware meta-learner utilizes the entropy values extracted from the entropy graph, to train our proposed model to recognize if an image pair is similar or dissimilar (i.e., belong to the same malware family or not)---see Algorithm \ref{algorithm:generator2}. To obtain an entropy graph, a malware binary file is read as a stream of bytes, and is separated into a few segments. The frequency of unique byte value is counted, and computes the entropy, using Shannon's formula, as follows:
\begin{equation}
		Ent=-\sum_{i}\sum_{j}M(i,j)logM(i,j)
		\label{eq:entropy}
\end{equation}
where $M$ is the probability of an occurrence of a byte value. The entropy obtains the minimum value of 0 when all the byte values in a binary file are the same, while the maximum entropy value of 8 is obtained when all the byte values are different. 
The entropy values are then represented as a stream of values that can be reshaped as an entropy graph. The entropy graph is then converted as a feature vector inputted through the convolutional extractor of a pre-trained network (e.g., VGG-16 \cite{simonyan2014very}). The summary of the steps involved in the entropy graph is described in Algorithm\ref{algorithm:generator2}:
\vspace{12pt}
	
\begin{algorithm}[H]
		\SetKwInOut{Input}{Input}
		\SetKwInOut{Output}{Output}
		
		\Input{$f$: malware binary file; $l$: segment length; $n$: the number of files }
		\Output{entropy graph matrix $m$}
		
		\While{not reach to n}{
			1. read $l$ bytes from $f$, and defined as segment $s$;
			
			2. \textbf{for} $i = 0$ to $255$ \textbf{do}
			
			3. \ \ \  compute the probability $p_i$ of $i$ appearing in $s$;\\
			
			4. \ \ \  compute the Shannon entropy
		}
		
		Generate entropy graph $m$\ \\
		\caption{Pseudo-code of entropy graph}
		\label{algorithm:generator2}
\end{algorithm}

\subsection{Weight generator via factorization}
	
	In the generic SNN approach, the feature extractor only uses the image feature: this approach, however, is no longer effective in the detection of obfuscated malware, where multiple obfuscated malware samples contain almost identical image features.
	
	
The problem is further complicated when only a few samples (i.e., less than five malware samples) exist (i.e., not enough malware feature information to use for classification, as very few variations of malware samples can be collected from a small number of malware samples). To address this issue, we present a new novel weight generation scheme based on the work presented by \cite{gidaris2018dynamic}. In our proposed model, the weight generator $G(.,.|\phi)$ receives, as input, the entropy vectors $\mathbf{W_{ave}}$ of a class, in addition to the image vectors $\mathbf{Z'} = {\{\mathbf{z^{'}_{i}\}}_{i=1}^N}^{'}$ of the $N'$ training samples of the corresponding class: this results in a weight vector $\mathbf{w'}=G(\mathbf{Z', W_{ave})}$.
	
In our model, the weight generator scheme is incorporated in the fully connected (FC) layer, to solve the non-linear issue that exists in the relationship between the entropy feature and the malware image. By integrating the weight generator into the FC layer, the weights of the features extracted before the FC layer can then be integrated better into calculating new and more optimized weights for the whole model. We generate weights by creating a weight combination. The weight combination produces the composite features that encode the non-linear connection in the feature space: this is done by multiplying the entropy features and image features together, such that the composite features learn a feature-embedding resistance to different obfuscated malware variations. Note that the dimension of the weight generator $g^i$ on the FC layer must be matched with the dimension of the weight size of the $ith$ feature layers in $F$, so that the weights $\mathbf{W_i} \in \mathbf{R}^{m \times n}$ can be decomposed using the following equation:

\begin{equation}
	 W^i = W^i_{sr} \cdot W^i_{ts}
	\label{eq:factorazation}
\end{equation}
where $\mathbf{W^i_{sr}}\in \mathbf{R}^{m \times n} $ is the average entropy feature vector for each malware family $\{{t}_{1},\ldots {t}_{N} \}$, and $\mathbf{W^i_{ts}} \in$ $\mathbf{R}^n$ is the task-specific image feature vector for each malware. With such factorization, the weight generators only need to generate the malware-specific parameters for each malware family in the lower dimension, and learn one set of parameters in the high dimension shared across all malware families.
	
\subsection{Loss Function}
Our proposed model uses two different types of loss functions. Embedding loss is used by our task-aware meta-learner network to compute a loss across the inter-class variance (e.g., the features in the feature embedding space of a CNN branch), while a hybrid loss is used by the differencing layer of the SNN, to compute the similarities across inter-class variants between an image pair.

\subsubsection{Embedding Loss for Meta-Learner}
The feature representation of the entropy graph of a malware class can be easily influenced by binary loss: this is because the use of binary loss can only give the probability of the distribution of distances between positive and negative image pairs, and cannot estimate the probabilities of distances between positive and negative image pairs across different malware variations, thus not being able to correctly classify similar pairs of images across obfuscated malware samples (i.e., not being able to learn a discriminative feature during the training procedure). To address this issue, we added a secondary cross-entropy loss, not only to learn the discriminative feature, but also to address the effect of overfitting caused by contrastive loss. This embedding loss is defined using the following equation:
\begin{equation}
L_{e} = -\frac{1}{N}\sum ^N_{i=1}\sum ^T_{t=1}log[\frac{exp(F(\mathbf{x_i};\theta_t))\cdot\mathbf{y_t^i}}{\sum^T_{j=1}exp(F(\mathbf{x_i};\theta_t))}] 
\end{equation}
where $\mathbf{x^i}$ represents the $i$th sample in the dataset of the size $N$. The one-shot encoding applied to the input based on the labels is indicated by $\mathbf{y_t}$ $\in$ $\{0, 1\}^t$, while $T$ indicates the number of tasks during training (e.g., either in the whole dataset or in the minibatch).	

\subsubsection{Hybrid Loss for Our SNN}
To calculate the similarity score for our proposed model, we propose a hybrid loss function comprised of a center loss and a constructive loss. The center loss proposed by Wen et al. \cite{wen2016discriminative} is a supplement loss function to the softmax loss for the classification task, which can learn to find a sample that can act as the center image of each class, and try to shorten the distance across the training samples of similar features by moving them to be as close to the center of the sample as possible. This center loss can be calculated as follows:
\begin{equation}
L_{c}=\frac{1}{2N}\sum_{N}^{i=1}\left \| \mathbf{d_w^{(i)}-c_{i}} \right \|_{2}^{2} 
\end{equation}

This approach, however, does not address the issue of moving apart from the training samples of dissimilar features. To address this issue, we propose the hybrid loss function integrated with the pairwise center, to better project the latent task embedding $e_t = T (t)$ into a joint embedding space that contains both the negative and positive center points. 
	
We have adopted a metric learning approach, where the corresponding learned feature is closer to the joint feature embedding for positive inputs of a given image pair, while the corresponding learned feature is far away from the joint feature embedding for negative inputs of a given image pair:
\begin{equation}
	\min L = {\beta} \min L_e+L_{b, c} 
\end{equation}
where $\beta$ is the hyperparameter to balance the two terms; in our study, we set it at 0.8. 

\section{Experiments}\label{sec:experiment}
In this section, we describe the details of the datasets we used for the experiments, the model configuration, and the results of our experiments. The results were obtained by running the experiments on a desktop with a  32 GM  RAM,  Nvidia Geforce RTX 2070(8GB),  and  Intel(R) Core(TM) i7-9700 CPU @ 3.00 GHz.

\subsection{Andro-Dumpsys Dataset}
We used the Andro-Dumpsys dataset obtained from \cite{Jang2016125}, which has been widely used for malware detection. The original dataset consists of 906 malicious binary files from 13 malware families. As illustrated in Table \ref{table:malware-family}, the number of malware variants and the total number of samples from different malware families varied. Almost half of the malware families had no more than 25 malware samples, while some only had 1 sample, as they were most likely the new malware detected lately (e.g., Blocal and Newbak). In addition to the original dataset, we also generated three additional synthetic malware variants, each of which was applied using the different control flow obfuscation techniques described earlier: function logic shuffling; junk code insertion; and function splitting, respectively. Two samples from each additional malware variant, a total of 6 additional samples for each malware family, were added to the original dataset.
\begin{table}[H] 
	\caption{Andro-Dumpsys Dataset with synthesis samples. \label{tab2}}
	\newcolumntype{C}{>{\centering\arraybackslash}X}
	\begin{tabularx}{\textwidth}{CCCC}
		\toprule
		\textbf{No.}& \textbf{Family} &\textbf{Number of Variants}   & \textbf{Number of Samples} \\
		& &  \textbf{(+3 Synthetic)}   & \textbf{(+ 6 Synthetic)}  \\
		\midrule
		1  & Agent	& 39 (42) &   150 (156)\\
		2&   Blocal  & 1 (4) &1 (7)\\
		3 & Climap & 1  (4)&  5 (11) \\
		4 & Fakeguard & 1 (4) &  10 (16)\\
		5 & Fech  & 1  (4)&3 (9)\\
		6& Gepew  & 4  (7) & 112 (118)\\
		7& Gidix  & 6 (9) & 108 (114)\\
		8  & Helir    &1 (4) &   15 (21)\\
		9&   Newbak  & 1  (4)  &1 (7)\\
		10 &Recal  & 2 (5) &  25 (31)\\
		11	&SmForw  & 23 (26) &  166 (172)\\
		12&   Tebak  & 10  (13)&93 (99)\\
		13& Wroba  & 23  (26) &108 (114)\\
		\bottomrule
	\end{tabularx}
	\label{table:malware-family}
\end{table}

\begin{figure}[H]
	\includegraphics[scale=0.6]{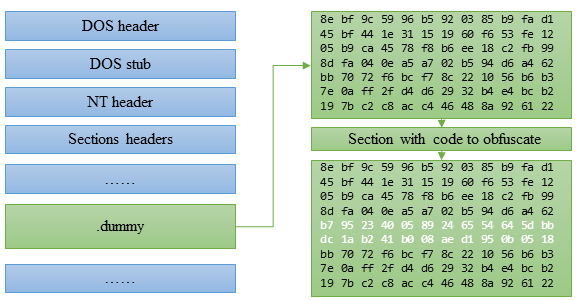}
	\caption{Example of inserting a junk code.}
	\label{fig:obf_code_example}
\end{figure}

Figure \ref{fig:obf_code_example} illustrates a snippet of how obfuscated malware is created by applying a junk code insertion. In this example, we created a dummy array that acts as a junk code, which we added in between two function calls from the original malware code. We applied a similar approach to the other two types of control flow obfuscation.

We also increased the image sample size, so as to have at least 30 samples for every malware family, using a data augmentation technique (e.g., applying random transformations, such as image rotations, re-scaling, and clipping the images horizontally).  The details of the augmentation parameters are shown in Table \ref{table:parameters}. In particular, ZCA whitening is an image preprocessing method that leads to transformation of the data, such that the covariance matrix is the identity matrix, leading to decorrelated features, while the fill\_mode argument with “wrap” simply replaces the empty area, and follows the filling scheme.
\begin{table}[H] 
	\caption{Parameters of data augmentation. \label{tab3}}
	\newcolumntype{C}{>{\centering\arraybackslash}X}
	\begin{tabularx}{\textwidth}{lll}
		\toprule
		\textbf{Parameters}	& \textbf{Values}	& \textbf{Description}\\
		\midrule
		rescale & 1./255  
		&   Resizing an image by a given scaling   factor. \\
		
		zca\_epsilon & $1e^{-6}$ 
		& Epsilon for ZCA whitening. \\
		fill\_mode & wrap & Points outside the boundaries of the input are filled accord-     \\
		&   & ing to the given mode.   \\
		rotation\_range & 0.1 &  Setting degree of range for random  rotations.\\
		
		height\_shift\_range & 0.5& Setting range for random vertical shifts.\\
		horizontal\_flip & True &  Randomly flips inputs  horizontally. \\
		\bottomrule
	\end{tabularx}
	\label{table:parameters}
\end{table}

\subsubsection{Image Feature}
We used the same technique proposed by \cite{zhu2020multi} to produce image features. To produce image features, we first read the binary malware as a vector of 8-bit unsigned integers, which were then converted into 2D vectors. We used the fixed width, while the height was decided according to the size of the original file. Finally, the 2D vector was converted into the gray images, using the color range [0, 255]. Note that the gray images at this stage were different dimensions according to the varying heights and widths in which size biases could occur in the fully connected layer: to address this issue, we used the bilinear interpolation method, as suggested by \cite{malvar2004high}, to produce our image feature in uniform to the size of 105$\times$105.

\subsubsection{Entropy Feature}
To obtain the entropy feature of each malware family, we took each byte of the malware binary file as a random variable, and counted the frequency of each value (00h-FFh). More concretely, the byte reads from the binary file were divided into several segments. For each segment, we first calculated the frequency of each byte value $p_j (1 \le j \le m)$, and then we calculated the entropy $y_i$ of the segment. 
The entropy values were then represented as a stream of values that could be reshaped as an entropy graph, with the size of  $254  \times 254 \times 1$. These entropy graphs were then converted as a 4096-dimensional feature vector inputted through the convolutional extractor of the VGG-16 architecture \cite{simonyan2014very}.

\subsection{Model Configurations}

\begin{figure}[H]
	\includegraphics[scale=0.55]{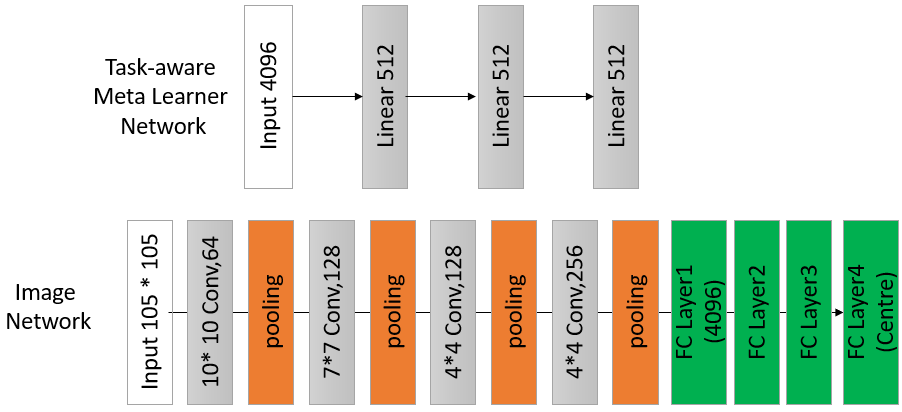}
	\caption{Network Configurations.}
	\label{fig:universe}
\end{figure}

The task-aware meta-learner network $\mathcal{T}$(t) was a two-layer FC network with a hidden unit size of 512, except for the top layer, which was 4096 for input. The weight generator $g^i$ was a single FC layer with the output dimension the same as the output dimension of the corresponding feature layer in $\mathcal{F}$. We added a ReLU function to the output of $g^i$ in cases where processed malware images were used as inputs, and the convolutional part was configured as the 4-layer convolutional network (excluding pooling layers), following the same structure as \cite{koch2015siamese}. In addition, the ReLU function and batch normalization were added afterwards by the convolution layer and the FC layer. The total number of parameters in our proposed model is 40 million. The overview of our network configurations is described in Figure \ref{fig:universe}.

\subsection{Results}
\begin{figure}[H]
	\includegraphics[scale=0.7]{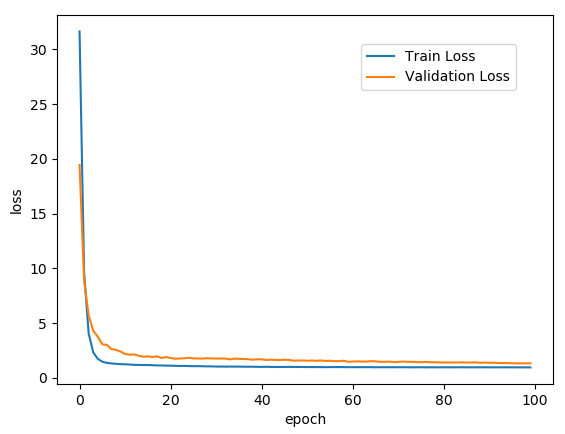}
	\caption{Loss during training on epochs.}
	\label{fig:convergence}
\end{figure}

We set the batch size to 32, and we used Adam as the optimizer, with an initial learning rate of $10^{-4}$ for the image network and the weight generators, and $10^{-5}$ for the task embedding network. The network was trained with 50 epochs, which ran for approximately 2 h. As Figure \ref{fig:convergence} illustrates, both train and validation loss stabilized after 50 epochs, confirming that the training had been done by this stage.

The testing process conducted M times of N-way on N-shot learning tasks, where $Q$ times of correct predictions contributed to the accuracy, calculated by the following formula:
\begin{equation}
	Accuracy=(100*Q/m)\% .
\end{equation}
\subsubsection{N-Way Matching Accuracy}
The evaluation of $N$-way learning at each test state was carried out for one-shot and five-shot. For $N$-way one-shot learning, we chose an anchor image from one class of test, and then randomly selected $N$ classes of images to form the support set $X = \{x_i\}^N_{i=1}$, where $x_1$, $\forall x \in X$; the selected image’s class was the same as the anchor image $\hat{x}$, and the other images in the support sets were from different classes. The similarity score between $\hat{x}$ and other images was calculated using our model. Specifically, if the similarity score of the feature vector of $x_1$, which could be represented as  $S = \{s_i\}^N_{i=1}$, was the maximum of $S$, then the task could be labeled as a correct prediction; otherwise, it was regarded as an incorrect prediction. For $N$-way five-shot learning,  we randomly selected $N$ unseen classes and six instances, in which five instances of each class were randomly selected as the support set, $X = \{x_1,\ldots ,x_i\}^N_{i=5}$, and the remaining instances of each class formed the query set: its prediction procedure was the same as the test in the one-shot.

\begin{figure}[H]
	\begin{tabular}{cc}
		
		\includegraphics[width=2.2in]{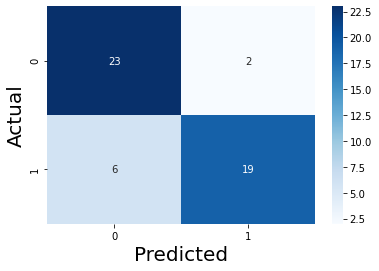}&
		\includegraphics[width=2.2in]{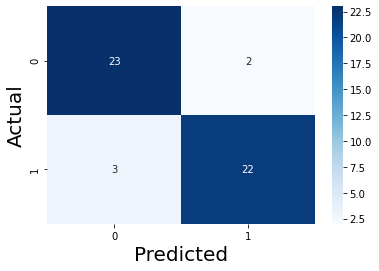}\\
		N-way one-shot&N-way five-shot
	\end{tabular}
	\caption{Matching Accuracy.}
	\label{fig:confusion}
\end{figure}

The matching accuracy of N-way accuracy for the one-shot and five-shot are illustrated in Figure~\ref{fig:confusion}. We randomly used 50 pairs of images, 25 containing positive image pairs and 25 containing negative image pairs, to test the effectiveness of our proposed model. As shown in the N-way one-shot result in Figure~\ref{fig:confusion}a, 19 out of 25 positive image pairs were matched correctly, while there were 6 true negatives (i.e., 2 positive pairs not matched correctly). Similarly, 23 of 25 negative pairs matched correctly, while there were 2 false positives (i.e., 2 negative pairs matched incorrectly). For the N-way five-shot results (shown in Figure~\ref{fig:confusion}b), the accuracy of matching was higher, as almost 22 out 25 pairs matched correctly for both positive and negative pairs, and there were 3 incorrectly matched results.

\begin{figure}[H]
	\subfloat[\centering Initial Stage]{%
		\includegraphics[width=.45\textwidth]{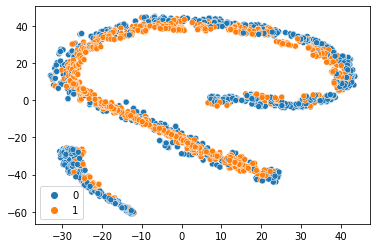}%
		\label{fig:y equals x}%
	}\hfill
	\subfloat[\centering Trained Stage]{%
		\includegraphics[width=.45\textwidth]{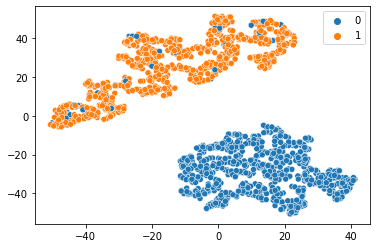}%
		\label{fig:y equals x}%
	}
	\caption{PCA visualization for classification performance.}
	\label{fig:epoch}
\end{figure}

Figure~\ref{fig:epoch} shows the projection of the embeddings space, using the two-dimensional principal component analysis (PCA) technique, where each orange point dictates the distance of a positive pair, while each blue point dictates the distance of a negative pair. Figure~\ref{fig:epoch}a shows the embedding space before training, while Figure~\ref{fig:epoch}b projects the embedding space after training. After training, we could clearly see two distinct clusters---one around the distance calculated for all positive pairs, and the other for negative pairs: this confirmed that our proposed model had learned well, so as to distinguish the similarities among positive pairs and negative pairs and separate them far apart.

\vspace{-17PT}

\subsubsection{Benchmark against similar methods}
Table \ref{tab:related work} shows the result of benchmarking our proposed model against the current state-of-the-art, especially the Matching network and Prototypical network, as well as the original Siamese network. Our model surpassed the performance of the Matching Network and Prototypical Network by $2.4\%$ and $1.8\% $ on the one-shot learning on the 5-way. The difference between our results of one-shot and five-shot was 3.1\%, 1.9\%, and 1.4\% on the 5-way, 10-way, and 15-way, respectively. Our proposed five-shot result outperformed all three exiting models by 1.4\%, 5.9\%, and 5.4\% on the 5-way, 10-way, and 15-way, respectively.
\begin{table}[H] 
	\caption{Comparison of classification performance of different few-shot learning approaches for Andro-Dumpsys datasets. \label{tab5}}
	\newcolumntype{C}{>{\centering\arraybackslash}X}
	\begin{tabularx}{\textwidth}{ClCCC}
		\toprule
		1-shot \\
		\cmidrule(l){1-2}
		Ref & \textbf{Method}	& \textbf{5-way}& \textbf{10-way}& \textbf{15-way}\\
		\midrule
		\cite{vinyals2016matching}   & Matching network  &  85 $\pm$ \ 2.2\% & 84 $\pm$ \ 2.4\% &  76 $\pm$ \ 2.6\%\\
		\cite{snell2017prototypical}  & Prototypical network   & 86 $\pm$ \ 1.7\%& 82 $\pm$ \ 1.7\%&81 $\pm$ \ 1.9\% \\ \cite{koch2015siamese}  & Siamese network & 82 $\pm$ \ 2.5\%& 69 $\pm$ \ 2.3\%& 64 $\pm$ \ 2.6\%  \\
		\ \    & Task-aware SNN  & 88 $\pm$ \ 2.2\%& 86 $\pm$ \ 2.2\%& 82 $\pm$ \ 2.4\%\\
		\toprule
		5-shot \\
		\cmidrule(l){1-2}
		Ref	& \textbf{Method}	& \textbf{5-way}& \textbf{10-way}& \textbf{15-way}\\
		\midrule
		\cite{vinyals2016matching}    & Matching network   &  89 $\pm$ \ 2.1\% & 86$\pm$ \ 2.1\% &  78$\pm$ \ 2.3\%   \\
		\cite{snell2017prototypical}       & Prototypical network     & 89 $\pm$ \ 1.2\%& 85$\pm$\ 1.4\% &82 $\pm$ \ 1.5\%  \\
		\cite{koch2015siamese}     & Siamese network & 85 $\pm$ \ 2.0\% & 72 $\pm$ \ 2.2\% & 69 $\pm$ \ 2.7\%    \\
		\ \     & Task-aware SNN  &91$\pm$ \ 1.8\%& 88$\pm$ \ 2.1\%&  83 $\pm$ \ 2.1\%   \\
		\bottomrule
	\end{tabularx}
	\label{tab:related work}
\end{table}

\subsubsection{Distance Measure Effectiveness}
\begin{figure}[H]
	\includegraphics[scale=0.7]{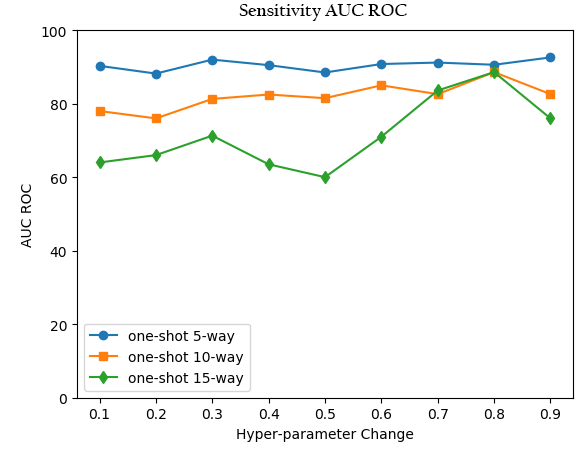}
	\caption{One-shot performance with hyperparameter changing.}
	\label{fig:hyper}
\end{figure}

\begin{figure}[H]
	\subfloat[\centering ROC of 5-way one-shot]{%
		\includegraphics[width=.45\textwidth]{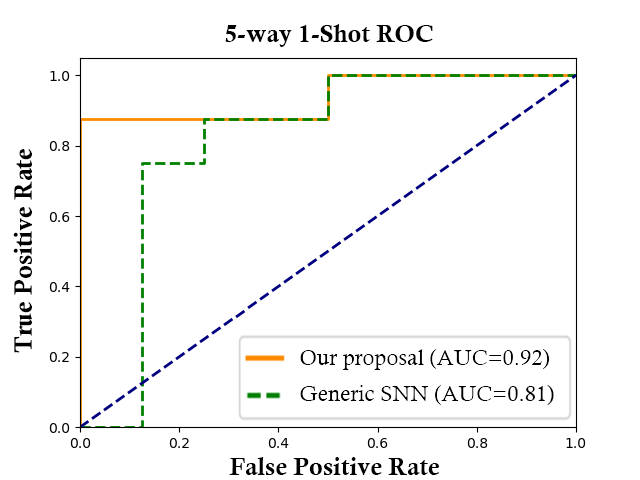}%
		\label{fig:y equals x}%
	}\hfill
	\subfloat[\centering ROC of 10-way one-shot]{%
		\includegraphics[width=.45\textwidth]{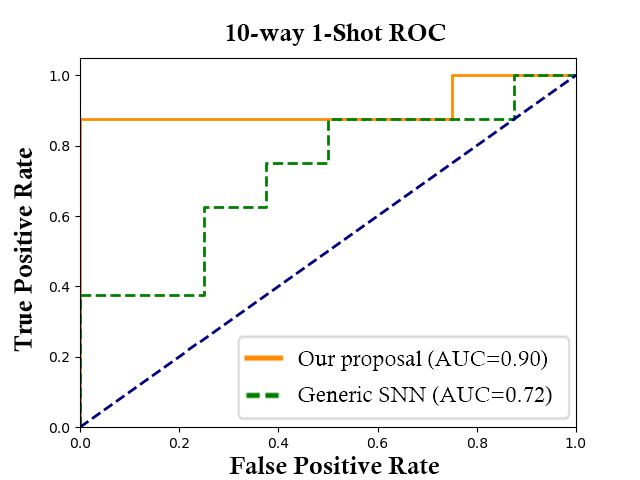}%
		\label{fig:y equals x}%
	}\hfill
	\subfloat[\centering ROC of 15-way one-shot]{%
		\includegraphics[width=.45\textwidth]{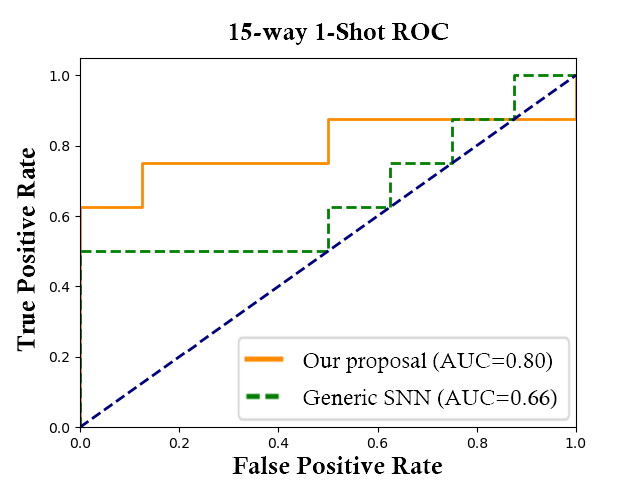}%
		\label{fig:y equals x}%
	}\hfill
	\subfloat[\centering ROC of 5-way five-shot]{%
		\includegraphics[width=.45\textwidth]{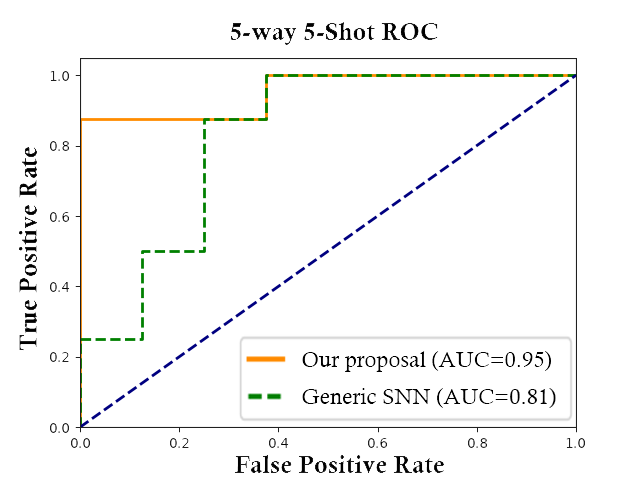}%
		\label{fig:y equals x}%
	}
	\hfill
	\subfloat[\centering ROC of 10-way five-shot]{%
		\includegraphics[width=.45\textwidth]{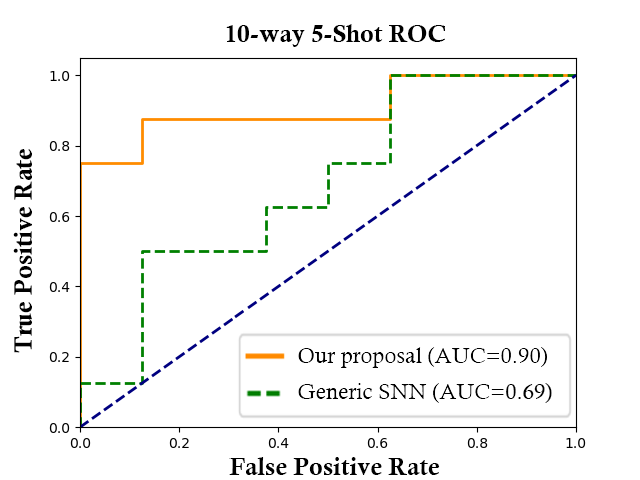}%
		\label{fig:y equals x}%
	}\hfill
	\subfloat[\centering ROC of 15-way five-shot]{%
		\includegraphics[width=.45\textwidth]{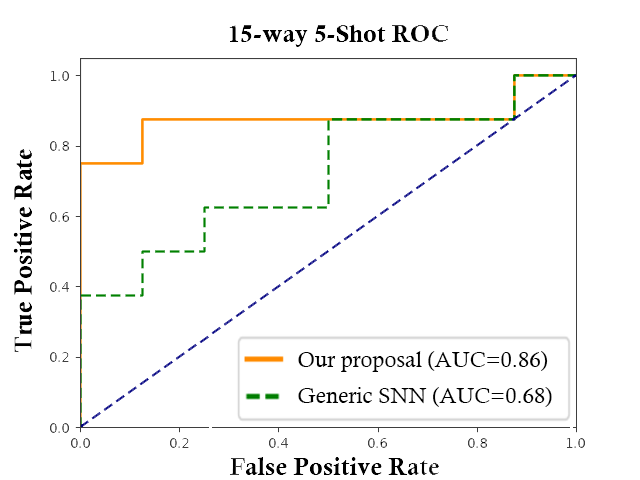}%
		\label{fig:y equals x}%
	}
	\caption{AUC--ROC curves under the N-way N-shot.}
	\label{fig:auc-roc}
\end{figure}

We also examined the effectiveness of our proposed model in distance measurement, by using the AUC (area under the curve) ROC (receiver operating characteristic) curve. The AUC--ROC curve is commonly composed of two performance measures: the true-positive (FPR) rate and the false-positive rate (FPR) rate. The equations related to these two performance measures of the AUC--ROC curve are shown as follows:
\begin{equation}
	\begin{aligned}
		FPR(P^*):=\int_{p^*}^1f_0(p)dp , \\
		TPR(P^*):=\int_{p^*}^1f_1(p)dp .
	\end{aligned}
\end{equation}
where the $f_0(p)$ is denoted by the probability of a density function for the predictions $p(x)$ produced by our proposed model. The negative pairs are labeled as 0, and $f_1(p)$ is the probability from the positive pair that is labeled as 1.  The given discrimination threshold $P$ is the integrals of the tails of these distributions according to the true-positive rate and false-positive rate. Based on the  two parameters TPR and FPR, the AUC--ROC curve is defined as follows:
\begin{equation} 
	AUC-ROC=\int_{0}^{1}TPR(FPR)D(FPR) .
\end{equation}
where the AUC measures the entire two-dimensional area underneath the entire ROC curve (i.e., integral calculus) from (0,0) to (1,1).  For example, a model whose predictions are 100\% wrong has an AUC of 0.0, while a model whose predictions are 100\% correct has an AUC of 1.0. Using this concept, we demonstrated the result predicted by the saved weights of one-shot and five-shot, respectively, of the learned model on the test set, including 5-way, 10-way, and 15-way. These AUC--ROC curves are shown in Figures \ref{fig:hyper} and \ref{fig:auc-roc}, which illustrate that the $\beta$ value at 0.8 provided the best performance when it was tested on one-shot N-way learning. 
This result also confirms that the hybrid multi-loss function can reduce the distance between the same classes, and enlarge the distance between different classes. Additionally, it does not change the attribute of the feature in the feature space, so the optimization of this layer will never negatively affect the deeper network layers.  This hybrid loss function can also compute classification accuracy with a learned distance threshold on distances.

As shown, our result is on a set of points in the true positive rate--false positive rate plane. The results achieved an AUC--ROC equal to 0.92, 0.91, and 0.80, respectively, under the 5-way, 10-way, and 15-way on the one-shot learning. We further conducted the AUC--ROC on the five-shot learning. Our proposed model also obtained better performance than the generic SNN, with 95.6, 90.7, and 86.8 \% at 5-way, 10-way, and 15-way, respectively. As expected, the accuracy of both one-shot and five-shot dropped as the number of N-way increased with higher intra-class variance.

Note that our model always performed better, as shown in these graphs, as the AUC--ROC areas (i.e., the areas up to the blue line) of our proposed model were larger compared to the generic SNN. 

\section{Conclusions}\label{sec:conclusion}
We propose a novel task-aware meta-learning-based Siamese neural network to accurately classify different malware families even in the presence of obfuscated malware variants. Each branch of the CNNs used by our model has an additional network called the “task-aware meta-learner network” that can generate task-specific weights, using the entropy graphs obtained from malware binary code. By combining the weight-specific parameters with the shared parameters, each CNN in our proposed model produces fully connected feature layers, so that the feature embedding in each CNN is accurately adjusted for different malware families, despite there being obfuscated malware variants in each malware family.

In addition, our proposed model can provide accurate similarity scores, even if it is trained with a limited number of samples. Our model also uses a pre-trained VGG-16 network in a meta-learning fashion, to compute accurate weight factors for entropy features. This meta-learning approach essentially solves the issues that are associated with creating potential bias due to not having enough training samples. 

Our model also offers two different types of innovative loss functions that can more accurately compute the similarity scores within a CNN and the feature embeddings used by two CNNs. 

Our experimental results show that our proposed model is highly effective in recognizing the presence of unique malware signatures, and is thus able to correctly classify obfuscated malware variants that belong to the same malware family. 

We are planning to apply different types of malware samples (e.g., DDoS attack \cite{wei2021ae} and ransomware families \cite{zhu2021task, zhu2021few, mcintosh2018large, mcintosh2019inadequacy}) and other data samples (e.g., finding similar abnormalities in medical X-ray images \cite{feng2022automated}), to test the generalizability of our model.

\section*{Acknowledgment}
This research is supported by the Cyber Security Research Programme—Artificial Intelligence for Automating Response to Threats from the Ministry of Business, Innovation, and Employment (MBIE) of New Zealand as a part of the Catalyst Strategy Funds under the
grant number MAUX1912.
\bibliographystyle{unsrtnat}
\bibliography{references.bib}  






\end{document}